\documentclass[runningheads]{llncs}
\usepackage{graphicx}
\usepackage{tikz}
\usepackage{comment}
\usepackage{amsmath,amssymb} % define this before the line numbering.
\usepackage{color}
\usepackage[pagebackref,breaklinks,colorlinks]{hyperref}

\usepackage[accsupp]{axessibility}  % Improves PDF readability for those with disabilities.

\usepackage{url}
\usepackage{cite}
\usepackage{multirow}
\usepackage{tabu}
\usepackage{booktabs}
\usepackage{mathrsfs}
\usepackage{bbm}

\begin{document}
\pagestyle{headings}
\mainmatter
\def\ECCVSubNumber{2953}  % Insert your submission number here

\title{Med-DANet: Dynamic Architecture Network for Efficient Medical Volumetric Segmentation} 
% Replace with your title

\titlerunning{Med-DANet}
\author{Wenxuan Wang\inst{1} \and
Chen Chen\inst{2} \and
Jing Wang\inst{1} \and
Sen Zha\inst{1} \and
Yan Zhang\inst{1} \and
Jiangyun Li\inst{1,\dagger}}
\authorrunning{W. Wang et al.}
\institute{
School of Automation and Electrical Engineering, University of Science and Technology Beijing, China,
\email{s20200579@xs.ustb.edu.cn, m202120718@xs.ustb.edu.cn, g20198675@xs.ustb.edu.cn, m202110578@xs.ustb.edu.cn, leejy@ustb.edu.cn}\\
\and
Center for Research in Computer Vision, University of Central Florida, USA, \email{chen.chen@crcv.ucf.edu}\\
\and
$\dagger$ Corresponding author: Jiangyun Li
}

%******************
\maketitle

\begin{abstract}

For 3D medical image (e.g. CT and MRI) segmentation, the difficulty of segmenting each slice in a clinical case varies greatly. Previous research on volumetric medical image segmentation in a slice-by-slice manner conventionally use the identical 2D deep neural network to segment all the slices of the same case, ignoring the data heterogeneity among image slices.
In this paper, we focus on multi-modal 3D MRI brain tumor segmentation and propose a dynamic architecture network named Med-DANet based on adaptive model selection to achieve effective accuracy and efficiency trade-off. For each slice of the input 3D MRI volume, our proposed method learns a \textit{slice-specific decision} by the Decision Network to dynamically select a suitable model from the predefined Model Bank for the subsequent 2D segmentation task. 
Extensive experimental results on both BraTS 2019 and 2020 datasets show that our proposed method achieves comparable or better results than previous state-of-the-art methods for 3D MRI brain tumor segmentation with much less model complexity. Compared with the state-of-the-art 3D method TransBTS, the proposed framework improves the model efficiency by up to 3.5$\times$ without sacrificing the accuracy. Our code will be publicly available at \url{https://github.com/Wenxuan-1119/Med-DANet}.

\keywords{Segmentation \and Brain Tumor \and MRI \and Dynamic Network \and Adaptive Inference}
\end{abstract}

\section{Introduction}
\label{sec:introduction}

Gliomas are the most common malignant brain tumors with different levels of aggressiveness. The precise measurements of gliomas can assist doctors in making accurate diagnosis and further treatment planning. Traditionally, the lesion regions are delineated by clinicians heavily relying on clinical experiences, which is time-consuming and prone to mistakes. Therefore, to improve the accuracy and efficiency of clinical diagnosis, automated and accurate segmentation of these malignancies on Magnetic Resonance Imaging (MRI) \cite{huo2017robust} is of vital importance.

In the past few years, deep neural networks, convolutional neural networks (CNNs) in particular, have achieved great success in medical image segmentation task.
The mainstream methods can be divided into two categories: (1) applying 2D networks for slice-wise (i.e. slice-by-slice) predictions and (2) utilizing 3D models (e.g. 3D CNNs) to process image volumes with multiple slices. 3D CNNs such as 3D U-Net \cite{3dunet} and V-Net \cite{vnet} employing 3D convolutions to capture the correlation between adjacent slices, have achieved impressive segmentation results. However, these 3D CNN architectures come with high computational overheads due to multiple layers of 3D convolutions, making them prohibitive for practical large-scale applications. Similarly, the 2D U-Net \cite{unet} and its variants such as \cite{zhang2018road,unet++,oktay2018attention} are also confronted with the same problem because of the unique architecture. Specifically, to obtain the multi-scale feature representation and fine-grained local details, multiple skip connections and stacked stacked convolutional layers are employed to improve model performance, but leading to unbearable computational overheads simultaneously.

Since the efficiency of a network determines the practical application value of the model deployment, model efficiency is as important as segmentation accuracy.
In order to cope with the high computational costs brought by 3D medical image itself and the segmentation networks mentioned above, many lightweight networks \cite{nuechterlein20183d,chen2018s3d,chen20193d,luo2020hdc,li2020memory,qin2021efficient} have been developed to realize efficient medical image segmentation. However, these proposed lightweight networks are designed from the perspective of efficient architecture without the consideration of data itself, treating all different inputs equally. Although these models effectively make the structural improvements to achieve lightweight architectures, they suffer from segmentation accuracy degradation due to reduced modeling capacity. Moreover, they can not \textbf{adaptively} make appropriate adjustments to different input data due to fixed network structure. Therefore, a natural question arises: 

\textit{For volumetric medical image segmentation task, is it possible to achieve dynamic inference with adjustable network structures for better accuracy and efficiency trade-offs by considering the characteristics of the input data (e.g. the level of segmentation difficulty of each image slice)?}

\begin{figure}[t]
    \centering
    \vspace{-5pt}
    \includegraphics[width=0.99\textwidth]{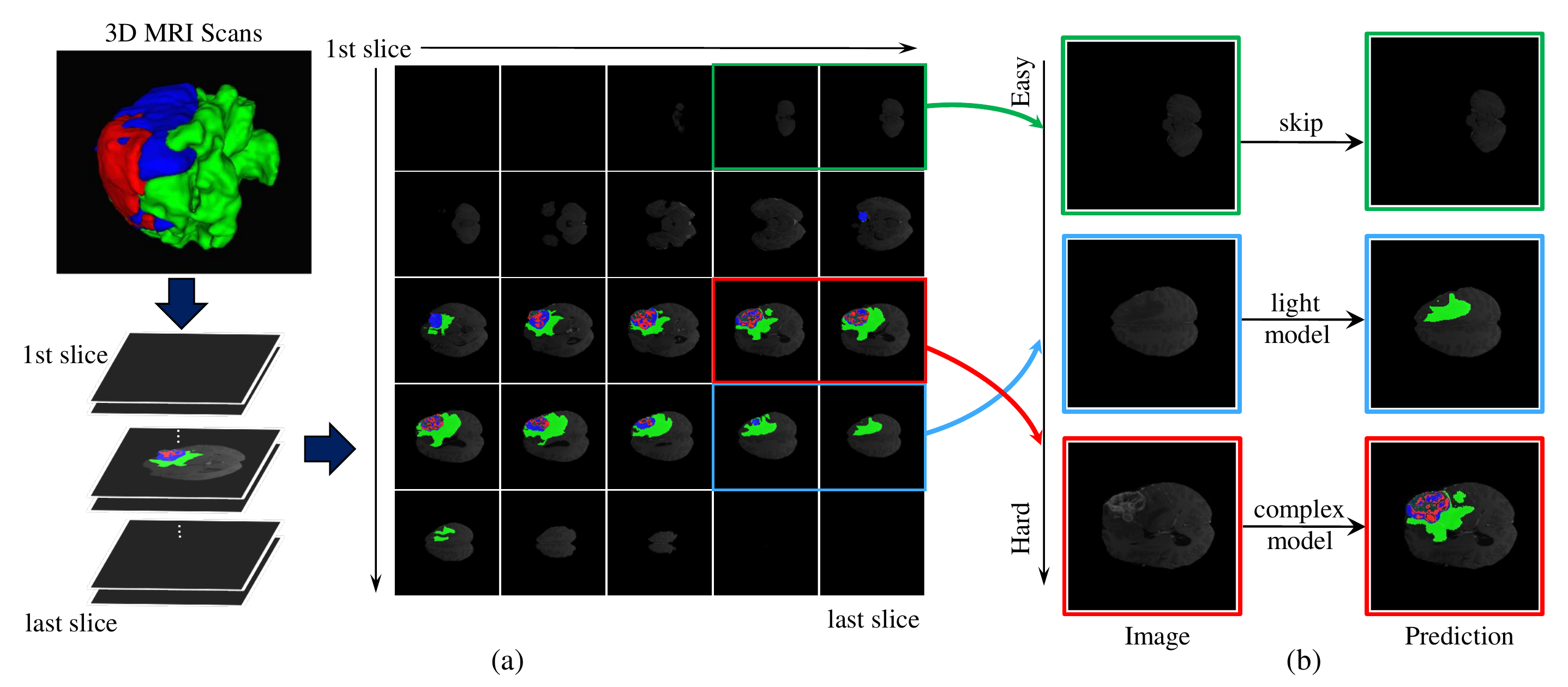}
    \caption{(a) The illustration of image content distribution along slice dimension of an MRI case (Axial View) from the BraTS 2019 dataset. The \textcolor{blue}{blue} regions denote the enhancing tumors, the \textcolor{red}{red} regions denote the non-enhancing tumors, and the \textcolor{green}{green} ones denote the peritumoral edema. (b) The main idea of our proposed framework for dynamic inference. For image slices with diverse segmentation difficulty, our framework realizes efficient and accurate segmentation by adaptively adjusting the architecture, selecting the optimal network in the Model Bank which consists of several networks with different model complexities. In this way, our framework can dynamically decide to ``\,slack off\," or ``\,work hard\," according to different samples.}
    \label{fig1}
\end{figure}

To answer this question, we take a brain tumor segmentation dataset BraTS 2019 \cite{menze2014multimodal,bakas2017advancing,bakas2018identifying} as an example to seek some insights. Fig. \ref{fig1} (a) shows the distribution of a 3D multi-modal brain tumor image along the slice dimension for one case. 
Due to several factors, such as the MRI process, shape of the organ (e.g. brain), and the location of the disease (e.g. glioma), the image content varies significantly across different MRI slices. For example, the $1^{st}$ row of Fig. \ref{fig1} (a) shows the first 5 slices that barely capture any tissue content of the brain. These slices can be simply predicted as containing all ``background" pixels (i.e. no lesion pixel) without model inference (i.e. ``skip" mode), saving the computational cost. For MRI slices contain lesions, the level of difficulty for segmentation also varies a lot. 
Some slices contain only certain categories of the foreground or tumor morphology is easy to segment (as highlighted with blue boxes in Fig. \ref{fig1} (b)), and some difficult-to-segment slices contain multiple types of tumors that are extremely irregular in shape and difficult to recognize (as highlighted with red boxes in Fig. \ref{fig1} (b)). From the analysis of the MRI data, the answer to the above question becomes clear -- Yes, it is possible to adjust the model complexity according to the input (e.g. image slice) for effective accuracy and efficiency trade-offs.

\textbf{Our Solution.} In this paper we tackle the aforementioned high computational overload problem of medical volumetric segmentation from a different perspective. Rather than designing more lightweight networks with static structure, we propose a highly efficient framework with \textbf{d}ynamic \textbf{a}rchitecture for \textbf{med}ical volumetric segmentation (Med-DANet). 
As illustrated in Fig. \ref{fig1} (b), taking a 2D image slice as input data, the Decision Network firstly generates a slice-dependent choice which represents the level of segmentation difficulty for the current slice. Then, according to the optimal choice made by the Decision Network, our method can adaptively determine to skip the current slice (i.e. directly output the segmentation map with only zero -- background class) as highlighted with green boxes in Fig. \ref{fig1} (b) or utilize the corresponding candidate segmentation network in the pre-defined Model Bank to accurately segment the current slice. The Model Bank consists of several networks with different model complexities. In this way, a reasonable allocation of computing resources for each slice is achieved by our adaptive segmentation framework.
The main contributions of this work can be summarized as follows:

\begin{itemize}
    \item 
    This work presents the \textit{first attempt} to explore the potential of dynamic inference in medical volumetric segmentation task. We focus on the 3D MRI brain tumor segmentation and propose a new framework with dynamic architectures to achieve a good balance between segmentation accuracy and efficiency. The proposed Med-DANet is generic and can be applied to any volumetric segmentation tasks (see \textcolor{blue}{Appendix} for the experiments of our Med-DANet on liver tumor segmentation with CT images).

    \item 
    By exploiting the special characteristics of multi-modal MRI brain tumor segmentation data that different slices have diverse degree of difficulty for segmentation, a comprehensive choice metric is designed to acquire the supervision signal for Decision Network, achieving the trade-off between accuracy and computational complexity of the model.
    
    \item
    Our proposed Med-DANet has strong scalability and flexibility. Any 2D networks can be incorporated into the Model Bank to meet various accuracy and efficiency requirements.
    
    \item   
    Extensive experiments on two benchmark datasets (BraTS 2019 and BraTS 2020) for multi-modal 3D MRI brain tumor segmentation demonstrate that our method reaches competitive or better performance than previous state-of-the-art methods with much less model complexity. 
\end{itemize}

\section{Related Work}

\subsection{Static and Lightweight CNNs for Medical Image Segmentation}
For medical image segmentation task, U-Net \cite{unet} and its variants \cite{3dunet,unet++,oktay2018attention} have achieved great success recently. However, the expensive computational costs impede the timely segmentation for assisting clinical diagnosis. To this end, great efforts have been made to design lightweight networks with improved model efficiency.
For example, 3D-ESPNet \cite{nuechterlein20183d} generalizes the efficient ESPNet \cite{mehta2018espnet} for 2D semantic segmentation to 3D medical volumetric segmentation, achieving satisfactory results on medical images. 
S3D-UNet \cite{chen2018s3d} takes advantages of the separable 3D convolution to improve model efficiency. DMFNet \cite{chen20193d} develops a novel 3D dilated multi-fiber network to bridge the gap between model efficiency and accuracy for 3D MRI brain tumor segmentation.  
HDCNet \cite{luo2020hdc} replaces 3D convolutions with a novel hierarchical
decoupled convolution (HDC) module to achieve a light-weight but efficient pseudo-3D model.
\cite{li2020memory} introduces a lightweight 3D U-Net with depth-wise separable convolution (DSC), which can not only avoid over fitting but also improve the generalization ability. 
In addition, knowledge distillation is also a popular method to achieve lightweight networks (i.e. student network). For example, \cite {qin2021efficient} proposes an efficient architecture by distilling knowledge from well-trained medical image segmentation networks to train another lightweight network for efficient medical image segmentation. 

\subsection{Dynamic Networks for Efficient Inference}

Lightweight models operates on the input data with the same static architecture, which cannot adaptively achieve the trade-off between accuracy and computational cost. To cope with this problem, dynamic networks are developed for efficient and adaptive inference \cite{yu2018slimmable,yang2020resolution,zhu2021dynamic,yang2020mutualnet,yang2021mutualnet,wang2021not}.
From the perspective of model architecture, the dynamic structure of network includes dynamic depth and dynamic width. For instance, slimmable networks \cite{yu2018slimmable} dynamically adjust the network width to achieve accuracy and efficient trade-offs at inference time. Moreover, adjusting input (e.g. image) resolution is also an effective way to balance between accuracy and efficiency. DRNet \cite{zhu2021dynamic} presents a novel dynamic-resolution network in which the resolution is determined dynamically based on each input sample. 

Apart from the research on dynamic inference mostly for the classification task, a few works aim to achieve dynamic inference in pixel labeling tasks. Kong et al. \cite{kong2019pixel} propose Pixel-wise Attention Gating to selectively process each pixel, allocating more computing power to pixels of fuzzy targets under specific resource constraints. Dynamic Multi-scale Network (DMN) \cite{he2019dynamic} adaptively learns weights of convolution kernels according to different input instances, arranging multiple DMN branches to learn multi-scale semantic information in parallel. Li et al. \cite{li2020learning}
introduce the concept of dynamic routing to generate data-dependent routes. Based on the scale distribution of objects in an image, the proposed soft condition gates can adaptively select scale transformation routes in an end-to-end manner.

\section{Methodology}
\label{method}
\subsection{Overview}

\begin{figure}
    \centering
    \vspace{-15pt}
    \includegraphics[width=0.99\textwidth]{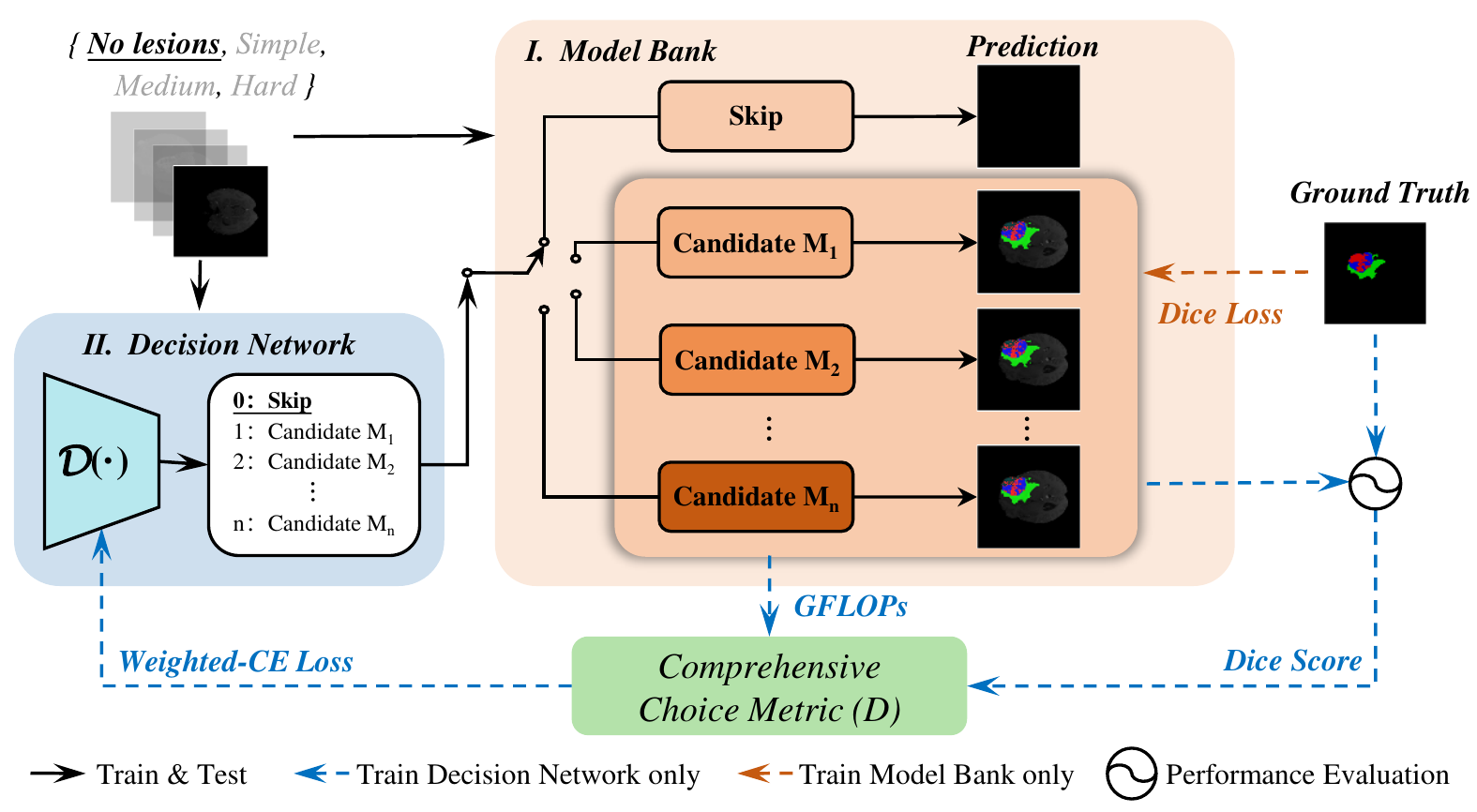}
    \caption{The illustration of the overall architecture of our proposed Med-DANet. Taking a 2D image slice as input data, the Decision Network generates a slice-dependent choice which represents the level of segmentation difficulties for the current slice. Then, according to the optimal choice made by the Decision Network, our method can adaptively determine to skip the current slice (i.e. directly output a segmentation map with only zero -- ``background" class) or utilize the corresponding candidate network in the pre-defined Model Bank (containing several networks with different model complexities) to accurately segment the current slice.}
    
    \label{fig2}
\end{figure}

An overview of our Med-DANet is shown in Fig. \ref{fig2}. In general, our framework consists of an extremely lightweight Decision Network ($\mathcal{D}$) and a Model Bank ($\mathcal{B}$) which contains $n$ different medical image segmentation networks (${M_1,M_2..., M_n}$). Models in the bank should be diverse in terms of the number of parameters and computational cost. 
To deal with the medical image datasets where segmentation targets are sparsely distributed among slices, we learn a \textit{slice-specific} decision by the Decision Network to dynamically select a suitable model from the Model Bank for the subsequent segmentation task, as formulated by Eq. \ref{eq1}. 
\begin{equation}
    \label{eq1}
    % y = \mathcal{D}(x)\circ\mathcal{B}(x),
    y = \mathcal{D}(x)\circ\mathcal{B}(x),
\end{equation}
where $x$ denotes the input image and $y$ is the corresponding prediction. $\mathcal{D} \circ \mathcal{B}$ indicates to take the matched element with index $\mathcal{D}$ in the collection $\mathcal{B}$, and the calculation details will be explained in the next subsection.

Roughly speaking, the Decision Network will comprehensively considers the segmentation accuracy and efficiency of each model, making the most appropriate choice. As for the Model Bank, any 2D networks can be included to meet various accuracy and efficiency requirements. More discussions on the model choices and ablation study are presented in Sec.~\ref{experimentalsetup} and Sec.~\ref{ablationstudies}.

\subsection{Dynamic Selection Policy}
We reduce the channel size of ShuffleNetV2 \cite{ma2018shufflenet} to get an extremely lightweight classification network as our Decision Network so that its computational overhead is negligible in the entire framework. The Decision Network undertakes a $n+1$-class classification task, and the $n+1$ categories refer to the $n$ candidate networks and a skip procedure. Therefore, the Decision Network and Model Bank can be respectively formulated as
\begin{equation}
    \label{eq2}
    \mathcal{D}(x)=\{\hat{D}|x;\theta\},
\end{equation}
\begin{equation}
    \label{eq3}
    \mathcal{B}(x)=[\varnothing,M_1(x),M_2(x),...,M_n(x)],
\end{equation}
here $\theta$ represents all the parameters of the Decision Network and $\hat{D}$ is the prediction of $\mathcal{D}(x)$. $M_1 \sim M_n$ denote the model candidates and $\varnothing$ indicates the skip procedure.

To be specific, when encountering a slice with only background (background slice, lesions are considered as foreground), the Decision Network will choose to directly skip the subsequent segmentation process. 
Otherwise, the Decision Network dynamically selects an appropriate segmentation model considering the recognition difficulty of foreground objects. 
During training process, the supervision of the Decision Network comes from the trade-off between model performance and efficiency. The calculation of the supervision signal of the decision process is as follows
\begin{equation}
    \label{eq4}
    D=\left\{
    \begin{array}{lr}
        0, & P_f < 1 \\
        argmax((1-\alpha)*S_i + \alpha*softmax(\frac{1}{F_i})) + 1, & P_f \geqslant 1
    \end{array},
    \right.
\end{equation}
where $S_i$ and $F_i$ is respectively the Dice Score and FLOPs of candidate model $M_i$ during the model training. $P_f$ denotes the number of foreground pixels (all pixels of segmentation targets). Specifically, if the number of foreground pixels is less than 1 (i.e. $P_f<1$), the current slice will be considered without any lesion areas, which should be directly skipped (i.e. the corresponding supervision is 0) during inference. Note that we normalize $\frac{1}{F_i}$ through the softmax operation to avoid the negative effects of the order of magnitude difference between accuracy and computations, in case that the acquired $D$ is dominated by either model performance or complexity. In addition, $\alpha$ is a coefficient to moderate the impact of Dice Score and FLOPs.

Given the choice $\hat{D}$ predicted by $\mathcal{D}(x)$ and the ground-truth $D$ calculated with Eq. \ref{eq4}, we apply the weighted cross-entropy loss to supervise our Decision Network. This allows the network to learn to skip (assigning all pixels directly to the background class without going through the segmentation models) the pure background slices in the dataset and comprehensively measure the accuracy ($S_i$) as well as efficiency ($\frac{1}{F_i}$) of different models for the segmentation targets from individual slice. In practice, the skip procedure is essential and can be widely applied because some background slices barely capturing any image content are very common in medical volumes, it is pointless to invest too much computation on these background slices. Moreover, the recognition difficulty of segmentation targets varies from slice to slice, it is more efficient to dynamically select segmentation models of different complexities.

\subsection{Training and Inference Strategy}
\textbf{Training}. The training process of the entire framework consists of two steps. First, the ensemble training of segmentation models. To save the training time cost, the $n$ segmentation models are jointly trained, minimizing the mean average of the dice-losses of all models and performing gradient back-propagation synchronously.
\begin{equation}
    \label{eq5}
    diceloss_j= \sum_{i=1}^{C}(1-\frac{2|pred_i \cap truth_i|}{|pred_i|+|truth_i|}),
\end{equation}
\begin{equation}
    \label{eq6}
    Loss_{\mathcal{B}}=\frac{1}{n}\sum_{j=1}^{n}diceloss_j.
\end{equation}
Here $C$ denotes the number of segmentation classes of the dataset, $diceloss_j$ is the dice-loss of candidate segmentation model $M_j (j\in\{1,2,...,n\})$ and $Loss_{\mathcal{B}}$ is the overall loss when training the Model Bank.

After that, we train the Decision Network based on a weighted cross-entropy loss:
\begin{equation}
    \label{eq7}
    Loss_{\mathcal{D}}=WCE(D,\hat{D})=-\sum_{i=0}^{n}w_i*d_i*log(\hat{d_i}),
\end{equation}
where WCE is short for weighted cross-entropy, $d_i$ and $\hat{d_i}$ are respectively the ground-truth and logits predicted by the Decision Network for model candidate $i$, $w_i$ represents the corresponding loss weight.

To cope with the problem of class imbalance (background slices make up a considerable portion of the dataset) and further pursue a better trade-off between segmentation accuracy and model complexity, we slightly enlarge the loss weights of candidate models with better performance (i.e. relatively lower the loss weight of the skip procedure). 

\noindent \textbf{Inference}. 
After the two-step training phase mentioned above, the well-trained decision network and predefined Model Bank are cascaded sequentially to achieve the final model structure at inference stage. Given a 2D slice as input image, our extremely lightweight Decision Network will decide to skip the current slice or choose the most appropriate segmentation network in the Model Bank based on the segmentation difficulty of the current slice. Following the specific selective choice made by Decision Network, the current slice will be directly skipped (i.e. output the corresponding segmentation maps with all zeros) or segmented by the single activated segmentation network included in the Model Bank. In this way, a dynamic slice-dependent framework with greatly improved efficiency is realized by our method. On one hand, compared with the previously proposed lightweight networks with static structure, our Med-DANet makes dynamic structure adjustments for different inputs instead of treating all inputs equally. On the other hand, compared with the previously proposed dynamic methods that utilize prediction confidence to determine whether the cascaded architecture need to early exit or not, our highly efficient Med-DANet can achieve the accurate segmentation in a one-pass manner.

\section{Experiments}

\subsection{Experimental Setup}
\label{experimentalsetup}

\textbf{Data and Evaluation Metric.}
The first 3D MRI dataset used in the experiments is provided by the Brain Tumor Segmentation (BraTS) 2019 challenge \cite{menze2014multimodal,bakas2017advancing,bakas2018identifying}. It comprises 335 patient cases for training and 125 cases for validation. Each sample is composed of 3D brain MRI scans with four modalities. Each modality has a volume of $240\times240\times155$ that has already been aligned into the same space. 
The ground truth include 4 classes: background (label 0), necrotic and non-enhancing tumor (label 1), peritumoral edema (label 2) and GD-enhancing tumor (label 4). 
The segmentation accuracy is measured by the Dice score and the Hausdorff distance (95\%) metrics for enhancing tumor region (ET, label 4), regions of the tumor core (TC, labels 1 and 4), and the whole tumor region (WT, labels 1,2 and 4), while the computational complexity is evaluated by the FLOPs metric.
The second 3D MRI dataset is provided by the Brain Tumor Segmentation Challenge (BraTS) 2020 \cite{menze2014multimodal,bakas2017advancing,bakas2018identifying}. It is comprised of 369 cases for training, 125 cases for validation. 
Except for the number of samples in the dataset, the other information about these two MRI datasets are identical.

\noindent \textbf{Implementation Details.}
The proposed Med-DANet is implemented on Pytorch \cite{paszke2019pytorch} and trained with 2 NVIDIA Geforce RTX 3090 GPUs (each has 24GB memory). For the \textbf{training} aspect, we first jointly train the Model Bank for 400 epochs from scratch with a batch size of 64. To prevent the small-scale candidates in the Model Bank from overfitting and make sure the large models can be fully optimized, we let small-scale candidates detach the training process at the epoch of 300 and make large-scale candidates continue to back propagate in the remaining epochs. After acquiring the training labels for Decision Network using our proposed comprehensive choice metric, the Decision Network is trained for 50 epochs from scratch with a batch size of 64. 
The Adam optimizer and the poly learning rate strategy with warm-up are utilized to train both two parts of our method. The initial learning rate for training the Decision Network and Model Bank are 0.01 and 0.0001, respectively.
Random cropping, random mirror flipping and random intensity shift are applied as the data augmentation techniques for training both the Decision Network and the segmentation candidates.
The softmax Dice loss and weighted cross-entropy loss are employed to train the Model Bank and the Decision Network respectively. Besides, $L2$ Norm is applied for model regularization with a weight decay rate of $10^{-5}$. 

As for the aspect of \textbf{model candidate selection} in the Model Bank, we choose the modified 2D UNet with various channel sizes (i.e. model width) and the 2D version of TransBTS \cite{wang2021transbts} with different scales (i.e. model depth) in this paper. The reason of choosing these two baselines is that both of them are state-of-the-art methods for brain tumor segmentation with excellent performance and they also represent two popular network architectures (i.e. CNN and vision transformer) that can extract complementary information from the data. Compared with the original UNet \cite{unet}, the modified version make improvements on both segmentation accuracy and efficiency. Taking consideration of both model depth and width, modified 2D UNet with a base channel of 12 (i.e. M1), modified 2D UNet with a base channel of 16 (i.e. M2), the light version of 2D TransBTS with 1-layer Transformer (i.e. M3), and 2D TransBTS with the original 4-layer Transformer (i.e. M4) are selected as the 4 model candidates in the Model Bank.
According to the policy made by the Decision Network, the modified 2D UNet can segment the easy slice with greatly reduced computations, while the 2D version of TransBTS achieves precise segmentation of the difficult slices by modeling explicit long-range dependency. In this way, with the well-trained Decision Network and the splendid segmentation candidates in Model Bank, our framework can achieve great trade-off between segmentation accuracy and efficiency. 

\subsection{Results and Analysis}

\textbf{BraTS 2019.} 
We conduct experiments on the BraTS 2019 validation set and compare our Med-DANet with previous state-of-the-art (SOTA) approaches.

\begin{table}[htpb]
\scriptsize
    \centering
    \caption{Performance comparison on BraTS 2019 validation set.} 
    \vspace{-5pt}
    \label{tab:comparison2019}
    {
    \begin{tabular}{c|c|c|c|c|c|c|c|c}
        \toprule[1.2pt]
        \multirow{2}{*}{Method} & \multicolumn{3}{c|}{Dice Score (\%) $\uparrow$} & \multicolumn{3}{c|}{Hausdorff Dist. (mm) $\downarrow$} &\multicolumn{2}{c}{FLOPs (G) $\downarrow$}\\
        \cline{2-9}
        &  ET &  WT &  TC &  ET &  WT & TC & per case & per slice\\
        \hline
        \centering
        3D U-Net \cite{3dunet}  & 70.86 & 87.38 & 72.48 & 5.062 & 9.432 & 8.719 & 1,669.53 & 13.04 \\
        V-Net  \cite{vnet}    & 73.89 & 88.73 & 76.56 & 6.131 & 6.256 & 8.705 & 749.29 & 5.85 \\
        Attention U-Net  \cite{oktay2018attention}   & 75.96 & 88.81 & 77.20 & 5.202 & 7.756 & 8.258 & 132.67 & 1.04 \\
        Wang et al. \cite{wang20193d}         & 73.70 & 89.40 & 80.70 & 5.994 & 5.677 & 7.357 & - & - \\
        Chen et al.  \cite{chen2019aggregating}   & 74.16 & \textbf{90.26} & 79.25 & 4.575 & \textbf{4.378} & 7.954 & - & - \\
        Li et al. \cite{li2019multi}           & 77.10 & 88.60 & 81.30 & 6.033 & 6.232 & 7.409 & -  & - \\
        Frey et al. \cite{frey2019memory}      & 78.70 & 89.60 & 80.00 & 6.005 & 8.171 & 8.241 & - & - \\
        TransUNet  \cite{chen2021transunet}   & 78.17 & 89.48 & 78.91 & 4.832 & 6.667 & 7.365 & 1205.76 & 9.42 \\
        Swin-UNet  \cite{cao2021swin}   & 78.49 & 89.38 & 78.75 & 6.925 & 7.505 & 9.260 & 250.88 & 1.96 \\
        TransBTS  \cite{wang2021transbts}   & 78.36 &  88.89 & \textbf{81.41} & 5.908 & 7.599 & 7.584 & 333.09 & 2.60 \\
        \hline
        \bf{Ours}          & \textbf{79.99} & 90.13 & 80.83 & \textbf{4.086} & 5.826 & \textbf{6.886} & \textbf{77.78} & \textbf{0.61}\\       
        \bottomrule[1.2pt]
    \end{tabular}
    }
    
\end{table}

The \textbf{quantitative results} are presented in Table \ref{tab:comparison2019}. Our method achieves the Dice scores of $79.99\%$, $90.13\%$, $80.83\%$ on ET, WT, TC, respectively,  which are comparable or higher results than previous SOTA methods presented in Table \ref{tab:comparison2019}. Besides, a considerable improvement has also been achieved for segmentation in terms of Hausdorff distance metric. It is worth noting that the model complexity of our Med-DANet is significantly less than other SOTA methods, while the segmentation performance of ours is extraordinary. For example, the computational complexity of TransBTS \cite{wang2021transbts} is \textbf{3.5} times that of the proposed Med-DANet, and the model computational complexity of TransUNet \cite{chen2021transunet} is surprisingly \textbf{15.4} times that of our method, which fully validates the effectiveness of adaptive architecture for dynamic inference.

\begin{figure}[!tp]
    \centering
    \vspace{-5pt}
    \includegraphics[width=0.90\textwidth]{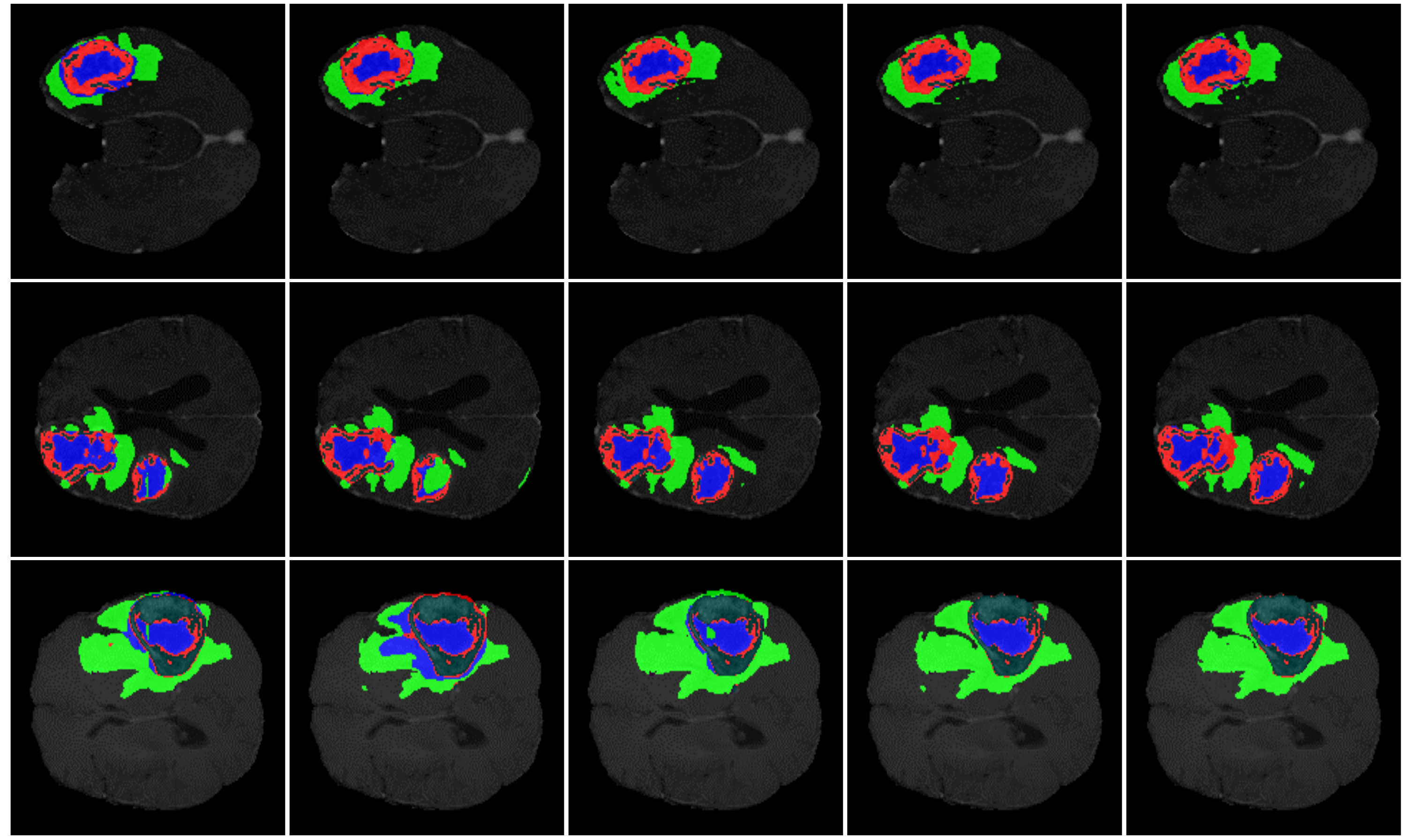}
    \begin{tabu} to 0.90\linewidth{X[1.0c] X[1.0c] X[1.0c] X[1.0c] X[1.0c]} 
        \scriptsize{3D U-Net} &  \scriptsize{VNet} &  \scriptsize{Att. U-Net} &  \scriptsize{\textbf{Ours}} &  \scriptsize{Ground Truth} \\
    \end{tabu}
    \caption{The visual comparison of MRI brain tumor segmentation results. The \textcolor{blue}{blue} regions denote the enhancing tumors, the \textcolor{red}{red} regions denote the non-enhancing tumors, and the \textcolor{green}{green} ones denote the peritumoral edema.}
    \label{fig3}
\end{figure}

For \textbf{qualitative analysis}, the brain tumor segmentation results of various methods are shown in Fig.~\ref{fig3} for a visual comparison (more visual comparison on BraTS 2019 dataset can be seen at Sec.~\ref{appendix1} in \textcolor{blue}{Appendix}), including 3D U-Net\cite{3dunet}, V-Net\cite{vnet}, Attention U-Net\cite{oktay2018attention}, and our Med-DANet. Since the labels for the validation set are not available, the five-fold cross-validation evaluation is conducted on the training set for all methods. 
It is obvious from Fig.~\ref{fig3} that our framework can delineate the brain tumors more accurately and generate much better segmentation masks with the powerful candidates as our dynamic options. Since we successfully take advantage of both CNNs and Transformer for different inputs, both local details and global context can be captured by our method to achieve accurate segmentation of tumors.

\noindent \textbf{BraTS 2020.} We also evaluate our Med-DANet on BraTS 2020 validation set and the segmentation results are reported in Table \ref{tab:comparison2020}. With the hyper-parameters on BraTS 2019 directly adopted for model training, our method achieves Dice scores of $80.57\%$, $90.28\%$, $81.34\%$ and HD of 6.474mm, 6.718mm, 7.416mm on ET, WT, TC. Considerable gain has been made by our method in terms of ET. Besides, compared with 3D U-Net \cite{3dunet}, V-Net \cite{vnet} and Residual 3D U-Net, our method makes great improvements in both metrics. It is clear that our method not only shows significant superiority in model performance but also has the great advantage of computational efficiency, which reveals the benefit of leveraging dynamic inference for medical volumetric segmentation task.

\begin{table}[htpb]
\scriptsize
    \centering
    \caption{Performance comparison on BraTS 2020 validation set. Per case and per slice denote the computational costs of segmenting a 3D patient case and a single 2D slice separately.}
    \label{tab:comparison2020}
    {
    \begin{tabular}{c|c|c|c|c|c|c|c|c}
        \toprule[1.2pt]
        \multirow{2}{*}{Method} & \multicolumn{3}{c|}{Dice Score (\%) $\uparrow$} & \multicolumn{3}{c|}{Hausdorff Dist. (mm) $\downarrow$} 
        &\multicolumn{2}{c}{FLOPs (G) $\downarrow$}\\
        \cline{2-9}
        &  ET &  WT &  TC &  ET &  WT &  TC & per case & per slice\\
        \hline
        \centering
        3D U-Net    \cite{3dunet}        & 68.76 & 84.11 & 79.06 & 50.983 & 13.366 & 13.607 & 1,669.53 & 13.04\\
        V-Net \cite{vnet}              & 61.79 & 84.63 & 75.26 & 47.702 & 20.407 & 12.175 & 749.29 & 5.85\\
        Deeper V-Net  \cite{vnet}              & 68.97 & 86.11 & 77.90 & 43.518 & 14.499 & 16.153 & - & -\\
        3D Residual U-Net \cite{zhang2018road}   & 71.63 & 82.46 & 76.47 & 37.422 & 12.337 & 13.105 & 407.37 & 3.18\\
        Liu et al. \cite{liu2020brain}   & 76.37 & 88.23 & 80.12 & 21.390 & 6.680 & \textbf{6.490} & - & -\\
        Vu et al.  \cite{vu2020multi}   & 77.17 & \textbf{90.55} & \textbf{82.67} & 27.040 & \textbf{4.990} & 8.630 & - & -\\
        Ghaffari et al. \cite{ghaffari2020brain}   & 78.00 & 90.00 & 82.00 & - & - & - & - & -\\
        TransUNet  \cite{chen2021transunet}   & 78.42 & 89.46 & 78.37 & 12.851 & 5.968 & 12.840 & 1205.76 & 9.42  \\
        Swin-UNet  \cite{cao2021swin}   & 78.95 & 89.34 & 77.60 & \textbf{11.005} & 7.855 & 14.594 & 250.88 & 1.96 \\
        TransBTS  \cite{wang2021transbts} & 78.50 & 89.00 & 81.36 & 16.716 & 6.469 & 10.468 & 333.09 & 2.60\\
        \hline
        \bf{Ours}          & \textbf{80.57} & 90.28 & 81.34 & \textbf{6.474} & 6.718 & 7.416 & \textbf{77.71} & \textbf{0.61}\\   
        \bottomrule[1.2pt]
    \end{tabular}
    }
\end{table}

\subsection{Ablation Studies}
\label{ablationstudies}

We conduct extensive ablation experiments to verify the effectiveness of our framework and justify the rationale of its design choices based on five-fold cross-validation evaluations on the BraTS 2019 training set.
(1) We make a fair comparison with each single candidate in the predefined Model Bank in terms of segmentation performance and computational cost.
(2) We investigate the effect of different designs for the final choice metric, which stands for the acquired supervision signal for Decision Network to help the proposed framework achieve optimal trade-off between accuracy and efficiency.
(3) We explore the effect of different lightweight networks for our Decision Network.
(4) We also analyze the effect of different numbers of candidate networks in the Model Bank. 
Besides, please check Sec.~\ref{appendix2} in \textcolor{blue}{Appendix} for more ablation study on BraTS 2019 training set.

\noindent \textbf{Comparison with Each Single Candidate in Model Bank.} We first compare our Med-DANet with all the candidates in Model Bank to demonstrate the powerful potential of dynamic architecture for medical volumetric segmentation. It is worth noting that the comparison is made under two different common settings to comprehensively evaluate the proposed framework. The first setting is utilize the cropped image with a spatial resolution of 128$\times$128 for training process and use slide-window technique to inference on original input with the spatial resolution of 240$\times$240 (i.e. full resolution), while the second setting is to utilize the full resolution for both training and inference. As presented in Table~\ref{tab:ablation1}, considerable improvements are achieved by our method in terms of both segmentation accuracy and model efficiency. Compared with the candidate M4 which has the largest model complexity under setting 1, our method achieves comparable performance with up to \textbf{8x} less computational costs. The same situation can be clearly seen under setting 2 in Table~\ref{tab:ablation1}. With much less model complexity and great segmentation performance, our proposed Med-DANet pursues the best trade-off between accuracy and efficiency, demonstrating the significance of adaptive architecture for dynamic inference.

\begin{table}[htpb]
\scriptsize
    \centering
    \caption{Comparison with each single candidate in the predefined Model Bank.} 
    \vspace{-5pt}
    \label{tab:ablation1}
    {
    \begin{tabular}{c|c|c|c|c|c|c|c}
        \midrule[1.2pt]
        \multicolumn{8}{c}{\textbf{Under Setting 1 (w/ slide window)}}\\
        \midrule[1.2pt]
        \multirow{2}{*}{Method} & \multicolumn{3}{c|}{Dice Score (\%) $\uparrow$} & \multicolumn{4}{c}{FLOPs (G) $\downarrow$} \\
        \cline{2-8}
        & ET & WT &  TC &  All Cases &  Per Case &  Per Slice &  Per Inference\\
        \hline
        \centering
        M1(Modified 2D UNet-12)           & 75.49  & 90.21  & 81.58  & \textbf{99,026.40}  & \textbf{1,500.40} & \textbf{9.68} & \textbf{0.61}\\
        M2(Modified 2D UNet-16)   & 78.31  & 90.61 & 82.59 & 174,646.57 & 2,646.16 & 17.07 & 1.07\\
        M3(2D TransBTS-light)    & 78.72  & 90.30 & 82.99 & 385,957.45 & 5,847.84 & 37.73 & 2.36\\
        M4(2D TransBTS)    & 77.21  & \textbf{91.08} & \textbf{83.27} & 814,962.73 & 12,347.92 & 79.66 & 4.98\\
        \hline
        \textbf{Ours}   & \textbf{78.75}  & 90.40 & 83.13 & 102,398.01 & 1,551.485 & 10.01 & 0.63\\
        \midrule[1.2pt]
        \multicolumn{8}{c}{\textbf{Under Setting 2 (w/o slide window)}} \\
        \midrule[1.2pt]
        \multirow{2}{*}{Method} & \multicolumn{3}{c|}{Dice Score (\%) $\uparrow$} & \multicolumn{4}{c}{FLOPs (G) $\downarrow$} \\
        \cline{2-8}
        &  ET &  WT & TC & All Cases & Per Case & Per Slice &  Per Inference\\
        \hline
        M1(Modified 2D UNet-12)    &  76.86 &  90.16 & 80.43  & \textbf{21,748.98} & \textbf{329.53} & \textbf{2.13}  & \textbf{2.13}\\
        M2(Modified 2D UNet-16)   & 77.22  & 89.99 & 81.61 & 38,362.50 & 581.25 & 3.75 & 3.75\\
        M3(2D TransBTS-light)  & 76.64  & 90.16 & 82.21 & 90,862.86 & 1,376.71 & 8.88 & 8.88\\
        M4(2D TransBTS)       &  76.48 & 90.57 & \textbf{83.21} & 203,351.94 & 3,081.09 & 19.88 & 19.88\\
        \hline
        \textbf{Ours}   & \textbf{77.73}  & \textbf{90.65} & 82.72 & 38,211.12 & 578.96 & 3.74 & 3.74\\
        \bottomrule[1.2pt]
    \end{tabular}
    }
\end{table}

\noindent \textbf{Effect of Different Designs for the Comprehensive Choice Metric.} To seek the optimal trade-off between model complexity and performance, we further investigate the effect of different designs for our proposed comprehensive choice metric (as illustrated in Eq. \ref{eq4}). As described in Sec.~\ref{method}, we introduce $\alpha$ and softmax operation to moderate the impact of Dice Score and FLOPs, in case that the acquired ground truth for Decision Network is dominated by either accuracy or complexity. The ablation results are listed in Table \ref{tab:ablation2}. It shows that $\alpha=0.001$ is the sweet spot for the whole framework to achieve the optimal balance between accuracy and efficiency.
Specifically, increasing $\alpha$ will make our method focus more on model efficiency, while the decreasing of $\alpha$ will push our method to pursue model accuracy without the consideration of computational cost.
Similarly, the drop of softmax operation on either Dice Scores or FLOPs will cause our framework to extremely pursue either the model performance or efficiency. 
By adopting the optimal configuration ($\alpha=0.001$, with softmax on FLOPs), our Med-DANet achieves greatly reduced computational complexity and competitive model accuracy.

\begin{table}

\scriptsize
    \centering
    \caption{Ablation study on effect of different design for the proposed comprehensive choice metric. ``$S$", ``$F$" denote the Dice Scores and FLOPs respectively, w/o and w/ denote with or without softmax on corresponding metrics (i.e. Dice Scores and FLOPs).} 
    \label{tab:ablation2}
    \vspace{-5pt}
    \resizebox{\linewidth}{!}
    {
    \begin{tabular}{c|c|c|c|c|c|c|c}
        \toprule[1.2pt]
        \multirow{2}{*}{Comprehensive Choice Metric Design} & \multicolumn{3}{c|}{Dice Score (\%) $\uparrow$} & \multicolumn{4}{c}{FLOPs (G) $\downarrow$} \\
        \cline{2-8}
        &  ET &  WT &  TC &  All Cases &  Per Case &  Per Slice &  Per Inference\\
        \hline
        \centering
        $\alpha$ = 0.0001   & \textbf{78.84}  & \textbf{90.48} & \textbf{83.28} & 129,768.88 & 1,966.20 & 12.69 & 0.79\\
        \centering
        $\alpha$ = 0.001  & 77.00  & 90.17 & 82.34 & 51,994.98 & 787.80 & 5.08 & 0.32\\
        \centering
         $\alpha$ = 0.01   & 77.21  & 90.19 & 81.80 & 49,951.03 & 756.83 & 4.88 & 0.31\\
        \hline
        $\alpha$ = 0.001, w/ S \& F & 77.00  & 90.17 & 82.34 & 51,994.98 & 787.80 & 5.08 & 0.32\\        
        $\alpha$ = 0.001, w/o F  & 75.57  & 90.10 & 81.85 & \textbf{48,255.83} & \textbf{731.15} & \textbf{4.72} & \textbf{0.29}\\
        \textbf{$\alpha$ = 0.001, w/o S}    & 78.75  & 90.40 & 83.13 & 102,398.01 & 1,551.485 & 10.01 & 0.63 \\
        $\alpha$ = 0.001, w/o S \& F    & 77.12  & 90.31 & 82.66 & 69,618.70 & 1,054.83 & 6.81 & 0.43 \\        
        \bottomrule[1.2pt]
    \end{tabular}
    }
    
\end{table}

\noindent \textbf{Effect of Different Lightweight Networks for Decision Network.} After investigating of the best design for the choice metric, we verify the effectiveness of our method with different Decision Networks. To achieve the highly efficient overall framework, the computational cost brought by the Decision Network should be controlled within acceptable limits. Therefore, four lightweight CNNs (MobileNetV2, GhostNet, ShuffleNetV2, and our modified ShuffleNetV2) are selected to study the influence of the Decision Network. To be noticed, the modified ShuffleNetV2 is acquired by greatly cutting down the channel size (i.e. model width). As shown in Table~\ref{tab:ablation3}, with our modified ShuffleNetV2 as the Decision Network, the proposed Med-DANet yields the best trade-off between accuracy and computational cost. Although our method achieves the best Dice Scores with MobileNetV2 as the Decision Network, the model complexity of MobileNetV2 and the overall FLOPs resulted by the guidance of MobileNetV2 are not acceptable. To be noticed, the model complexity of the modified ShuffleNetV2 is approximately \textbf{1/3} of GhostNet or ShuffleNetV2 and nearly \textbf{1/23} of MobileNetV2, which shows the effectiveness and efficiency of our optimal Decision Network. It is clear that employing modified ShuffleNetV2 enables our framework to show great superiority in terms of computation with competitive model performance.

\begin{table}[htpb]
\scriptsize
    \caption{Ablation study on effect of different choices for Decision Network. DN denotes the Decision Network, while ShuffleNetV2-M denotes our modified ShuffleNetV2.}
    \vspace{-10pt}
    \label{tab:ablation3}
    \centering
    \resizebox{\linewidth}{!}
    {
    \begin{tabular}{c|c|c|c|c|c|c|c|c}
        \toprule[1.2pt]
        \multirow{2}{*}{Decision Network} & \multirow{2}{*}{DN's FLOPs (G)} & \multicolumn{3}{c|}{Dice Score (\%) $\uparrow$} & \multicolumn{4}{c}{Overall FLOPs (G) $\downarrow$} \\
        \cline{3-9}
        &  &  ET &  WT &  TC &  All Cases &  Per Case &  Per Slice &  Per Inference\\
        \hline
        \centering
        MobileNetV2   & 1.758 & \textbf{78.54}  & \textbf{90.22} & \textbf{82.44} & 334,046.58 & 5,061.31 & 32.65 & 2.04\\
        GhostNet   & 0.278 &  75.57 & 90.19 & 81.48 & 82,536.60 & 1,250.56 & 8.07 & 0.50 \\
        ShuffleNetV2   & 0.247 & 77.20  & 90.19 & 81.48 & 96,606.43 & 1,463.73 & 9.44 & 0.59 \\
        \textbf{ShuffleNetV2-M} & 0.078 & 77.00  & 90.17 & 82.34 & \textbf{51,994.98} & \textbf{787.80} & \textbf{5.08} & \textbf{0.32}\\
        \bottomrule[1.2pt]
    \end{tabular}
    }
\end{table}

\noindent \textbf{Effect of Different numbers of Candidate Networks in Model Bank.} Finally, we conduct experiments to investigate the influence of the number of candidates in Model Bank on segmentation performance and efficiency. The quantitative results are illustrated in Table~\ref{tab:ablation4}. First of all, with no candidates (only skip procedure), the framework will naturally not work at all. Then we add the lightest CNN and Transformer as candidates ($n=2$), a good result has been achieved already. After that, the largest CNN and Transformer are also incorporated to Model Bank ($n=4$), making the segmentation performance and efficiency of the network both improve. Compared to 2 candidates, 4 candidates give the network more options for pursuing either performance or efficiency. However, when we further add 2 medium-sized CNN and Transformer to the Model Bank ($n=6$), although the network performance (i.e. segmentation accuracy) is further improved because of more optional network candidates, the computational cost is also increased. Moreover, more candidate networks in the model bank would also increase the training cost.
If higher precision requirements is necessary for the segmentation tasks, more candidates can be plugged into the Model bank to further boost the final performance. In conclusion, 4 candidates in Model Bank achieve the best balance between the accuracy and efficiency.

\begin{table}

\scriptsize
    \centering
    \caption{Ablation study on effect of different numbers of candidate networks ($n$) in Model Bank.}
    \vspace{-5pt}
    \label{tab:ablation4}
    {
    \begin{tabular}{c|c|c|c|c|c|c|c}
        \toprule[1.2pt]
        \multirow{2}{*}{number of candidate networks} & \multicolumn{3}{c|}{Dice Score (\%) $\uparrow$} & \multicolumn{4}{c}{FLOPs (G) $\downarrow$} \\
        \cline{2-8}
        &  ET &  WT &  TC &  All Cases &  Per Case &  Per Slice &  Per Inference\\
        \hline
        \centering
        0 (only skip)   & 9.09  & 0.00 & 0.00 & 0.00 & 0.00 & 0.00 & 0.00\\
        2 (1 CNN + 1 TR)    & 75.60 & 90.15 & 82.11 & 54,247.84 & 821.94 & 5.30 & 0.33\\
        \textbf{4 (2 CNN + 2 TR)}    & 77.00  & 90.17 & 82.34 & \textbf{51,994.98} & \textbf{787.80} & \textbf{5.08} & \textbf{0.32}\\
        6 (3 CNN + 3 TR)    & \textbf{78.71} & \textbf{90.76} & \textbf{82.92} & 84,130.02 & 1,274.70 & 8.22 & 0.51\\
        \bottomrule[1.2pt]
    \end{tabular}
    %}
    }
    
\end{table}

\vspace{-0.5cm}
\section{Conclusion and Discussion}

We present the \textit{first attempt} to explore the potential of dynamic inference in medical volumetric segmentation task. 
We focus on the 3D MRI brain tumor segmentation and propose a new framework named Med-DANet with dynamic architectures to achieve the trade-off between segmentation accuracy and efficiency. 
The proposed Med-DANet is generic and not limited to MRI brain tumor segmentation, which can be applied to any volumetric segmentation tasks (see Sec.~\ref{appendix3} in \textcolor{blue}{Appendix} for the experiments of our Med-DANet on liver tumor segmentation with CT images).
It is also worth noting that our proposed Med-DANet has strong scalability and flexibility. Any 2D state-of-the-art methods can be incorporated into our framework to satisfy different accuracy and efficiency requirements.
Extensive experiments on two benchmark datasets (BraTS 2019 and BraTS 2020) for multi-modal 3D MRI brain tumor segmentation demonstrate that our Med-DANet reaches competitive or better performance than previous state-of-the-art methods with greatly improved model complexity. 

\noindent \textbf{Broader Impact and Limitation.} Our approach provides a novel solution to efficient volumetric segmentation for medical applications, which inspires new research in this direction. Moreover, it is generalizable to other volumetric data (e.g. CT), please check Sec.~\ref{appendix3} in \textcolor{blue}{Appendix} for more details. One potential limitation could be the increased training cost due to several networks in the Model Bank, as compared to single network training. However, our joint training strategy and the multi-GPU training paradigm can greatly alleviate this issue. It also provides a future research direction to develop more efficient training schemes to match the cost of single network training. 

\section*{Acknowledgment}
This work was supported by the Fundamental Research Funds for the China Central Universities of USTB (FRF-DF-19-002), Scientific and Technological Innovation Foundation of Shunde Graduate School, USTB (BK20BE014).

\section{Appendix}
\label{appendix}

In this Appendix, we provide the following items:
\begin{enumerate}
    \item Experiments on the LiTS 2017 dataset \cite{bilic2019liver} (Liver Tumor Segmentation using \textbf{CT} scans), \textit{revealing the generalization ability of our proposed dynamic architecture framework (i.e. Med-DANet) on other medical imaging modalities (e.g. Computed Tomography (CT)) for the segmentation task.}
    \item More ablation study and analysis on BraTS 2019 and 2020 datasets for a comprehensive investigation.
    \item More visual comparison of brain tumor segmentation for qualitative analysis.
\end{enumerate}

\subsection{Experimental Results on LiTS 2017}
\label{appendix1}

\begin{figure}[t]
    \centering
    \vspace{-5pt}
    \includegraphics[width=0.99\textwidth]{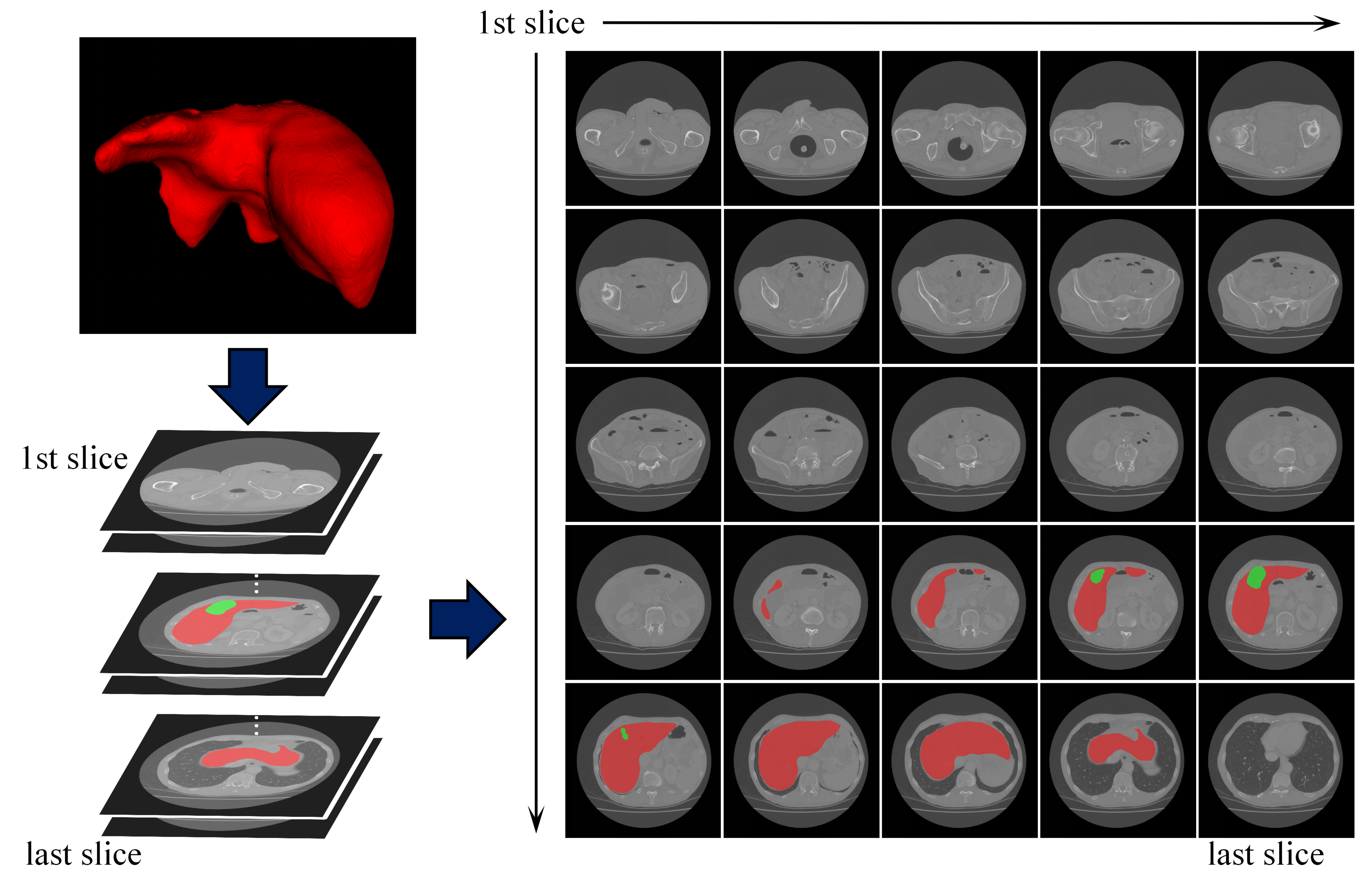}
    \caption{The illustration of image content distribution along slice dimension of a CT case (Axial View) from the LiTS 2017 dataset \cite{bilic2019liver}. The \textcolor{red}{red} regions denote the liver and the \textcolor{green}{green} regions denote the tumors.}
    \label{fig_supplementary_lits2017_singlecase}
    \vspace{-5pt}
\end{figure}

In order to show that our proposed insight is not just limited to segmentation task for MRI brain tumors but is a common phenomenon in different medial image modalities, we also present the image content distribution along the slice dimension of a 3D \textbf{CT} case from the LiTS 2017 dataset \cite{bilic2019liver} in Fig. \ref{fig_supplementary_lits2017_singlecase}. CT is another widely used imaging modality for various medical applications. Obviously, Fig. \ref{fig_supplementary_lits2017_singlecase} shows that the image content also varies significantly across different CT slices, which is similar to the distribution across diverse MRI slices. It is also evident that the segmentation difficulties are different among CT slices. Therefore, it is reasonable to adjust the model complexity according to different inputs (e.g. image slices) for effective accuracy and efficiency trade-offs.

\begin{table}[htpb]
\scriptsize
    \centering
    \caption{Performance comparison on LiTS 2017 testing set. ``P" refers to pre-trained model. Per case and per slice denote the computational cost of segmenting a 3D patient case and a single 2D slice, respectively.} 
    \vspace{-5pt}
    \label{tab:comparison_lits2017}
    {
    \setlength{\tabcolsep}{0.6mm}{
    \begin{tabular}{c|c|c|c|c|c|c}
        \toprule[1.2pt]
        \multirow{2}{*}{Method} & \multicolumn{2}{c|}{Dice per case (\%) $\uparrow$} & \multicolumn{2}{c|}{Dice global (\%) $\uparrow$} & \multicolumn{2}{c}{FLOPs (G) $\downarrow$} \\
        \cline{2-7}
        & Lesion & Liver  & Lesion & Liver  & Per Case & Per Slice\\
        \hline
        \centering
        U-Net \cite{chlebus2017neural}  & 65.00 & - & - & - & - & -\\
        3D DenseUNet w/o P \cite{li2018h}   & 59.40 & 93.60 & 78.80  & 92.90 & - & -\\
        2D DenseUNet w/o P  \cite{li2018h}   & 67.70 & 94.70 & 80.10  & 94.70 & - & -\\
        2D DenseNet w/ P  \cite{li2018h}   & 68.30 &95.30 & 81.80  & 95.90 & - & -\\
        2D DenseUNet w/ P  \cite{li2018h}   & 70.20 & 95.80 & \textbf{82.10}  & 96.30 & - & -\\
        I3D  \cite{carreira2017quo}   & 62.40 & 95.70 & 77.60  & 96.00 & - & -\\
        I3D w/ P  \cite{carreira2017quo}   & 66.60 & 95.60 & 79.90  & 96.20 & - & -\\
        Han \cite{han2017automatic}           & 67.00 & - & - & - & - & -\\
        Vorontsov et al. \cite{vorontsov2018liver}         & 65.00 & - & - & - & - & -\\
        TransUNet  \cite{chen2021transunet}   & 61.70 & 95.40 & 77.40 & 95.60  & 1200.64 & 9.38 \\
        Swin-UNet  \cite{cao2021swin}   & - & 92.70 & 67.60 & 91.60 & 249.60 & 1.95 \\
        TransBTS  \cite{wang2021transbts} & 70.30 & 96.00 & 81.50 & 96.40 & 330.00 & 2.58\\
        \hline
        \bf{Ours}  & \textbf{70.50} & \textbf{96.10}  & 81.90 & \textbf{96.60} & \textbf{37.12} & \textbf{0.29}\\
        \bottomrule[1.2pt]
    \end{tabular}
    \vspace{-10pt}
    }
    }
    
\end{table}

To evaluate the generalization ability of our proposed Med-DANet, we conduct experiments of liver tumor segmentation on {CT} scans using the LiTS 2017 dataset \cite{bilic2019liver}. 

The quantitative results on LiTS 2017 testing set are  presented in Table \ref{tab:comparison_lits2017}. It can be clearly seen that our method achieves comparable or even higher Dice scores than previous state-of-the-art methods with much less model complexity. Note that most of the comparison methods didn't provide the source codes. Therefore, we can not obtain the computational costs of those methods. In comparison with recently proposed Transformer based method named TransBTS \cite{wang2021transbts}) for medical image segmentation task (the source code of TransBTS is publicly available), our Med-DANet considerably advances the segmentation accuracy with greatly reduced computational costs. Specifically, the computational complexity of TransBTS \cite{wang2021transbts} is \textbf{8.89} times that of our Med-DANet, which is similar to the situation on BraTS 2019 and BraTS 2020 datasets. Thus, the results confirm the generalization ability and effectiveness of our adaptive framework with dynamic architecture.

\subsection{More Ablation Study and Analysis}
\label{appendix2}

In this section, to further explore the potential of our dynamic framework and justify the rationale of its design choices, more ablation experiments are conducted.
(1) We investigate the effect of different training strategies for training the model candidates in our Model Bank. Experiments are carried out using five-fold cross-validation evaluations on the BraTS 2019 training set.
(2) We present the selection ratio of each candidate model in the Model Bank for the BraTS 2019 and 2020 datasets.

\subsubsection{Effect of Different Training Strategies for Candidate Networks}

In order to employ the most suitable and efficient training approach for our Med-DANet, we investigate different strategies to train the candidate models in the model bank of our proposed framework. Since the Model Bank is composed of four candidate networks, the simple and straightforward way of training these candidates would be training each candidate individually (i.e. individual training). However, this individual training scheme is time-consuming. Therefore, in our proposed Med-DANet, we simultaneously train all the candidate networks together in a joint fashion (i.e. joint training). The comparison of the segmentation performance and training time for these two training strategies is shown in Table~\ref{tab:ablation1_supplementary}. The joint training method can greatly reduce the training time (i.e. up to \textbf{7.36 hours}) while achieving higher model accuracy. However, the individual training scheme yields better computational efficiency in terms of FLOPs.

\begin{table}[htpb]
\scriptsize
    \caption{Ablation study on effect of different training strategies for candidate networks.}
    \label{tab:ablation1_supplementary}
    \centering
    \begin{tabular}{c|c|c|c|c|c|c}
        \toprule[1.2pt]
        \multirow{2}{*}{Training Strategy} & \multicolumn{3}{c|}{Dice Score (\%) $\uparrow$} &\multicolumn{2}{c|}{FLOPs (G) $\downarrow$} & \multirow{2}{*}{Training Time (hour) $\downarrow$} \\
        \cline{2-6}
        & ET & WT & TC & Per Case & Per Slice \\
        \hline
        \centering
        Med-DANet(Individual Training)   & 75.73 & 90.25 & 82.31 & \textbf{962.87} & \textbf{7.52} & 21.99 \\
        Med-DANet(\textbf{Joint Training})    & \textbf{78.75}  & \textbf{90.40} & \textbf{83.13} & 1,551.485 & 10.01 & \textbf{14.63}\\
        \bottomrule[1.2pt]
    \end{tabular}
    \vspace{-20pt}
\end{table}

\subsubsection{Activation Ratio of Each Candidate Network}
 
\begin{figure}[!tp]
    \centering
    \includegraphics[width=0.95\textwidth]{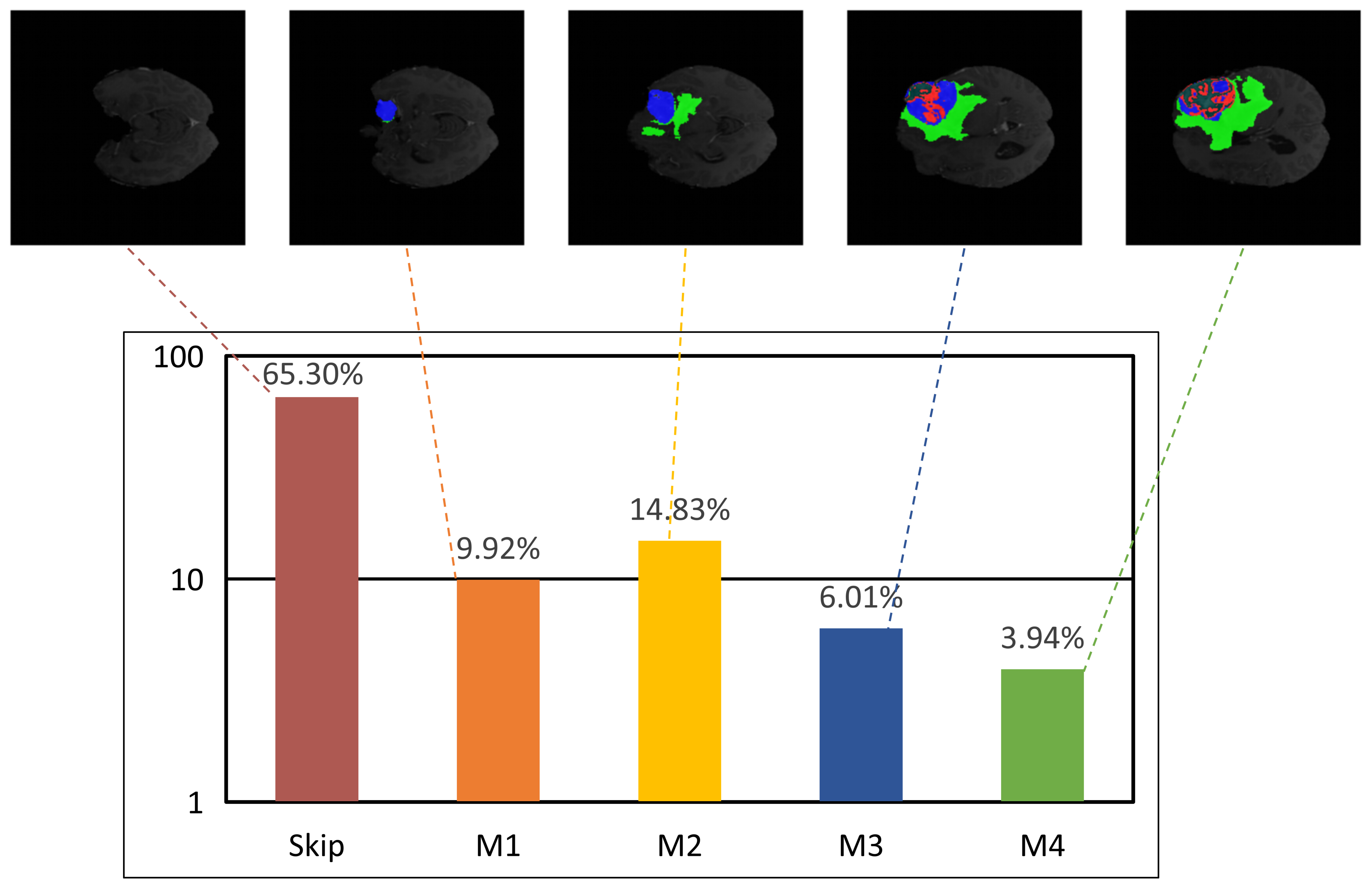}
    \caption{The activation ratio of each candidate model for different medical image slices in BraTS 2019 dataset. Skip, M1, M2, M3, M4 denote the operation of directly skip, candidate 1, candidate 2, candidate 3, and candidate 4, respectively.}
    \label{fig_supplementary_activation_brats2019}
\end{figure}

\begin{figure}[!tp]
    \centering
    \includegraphics[width=0.95\textwidth]{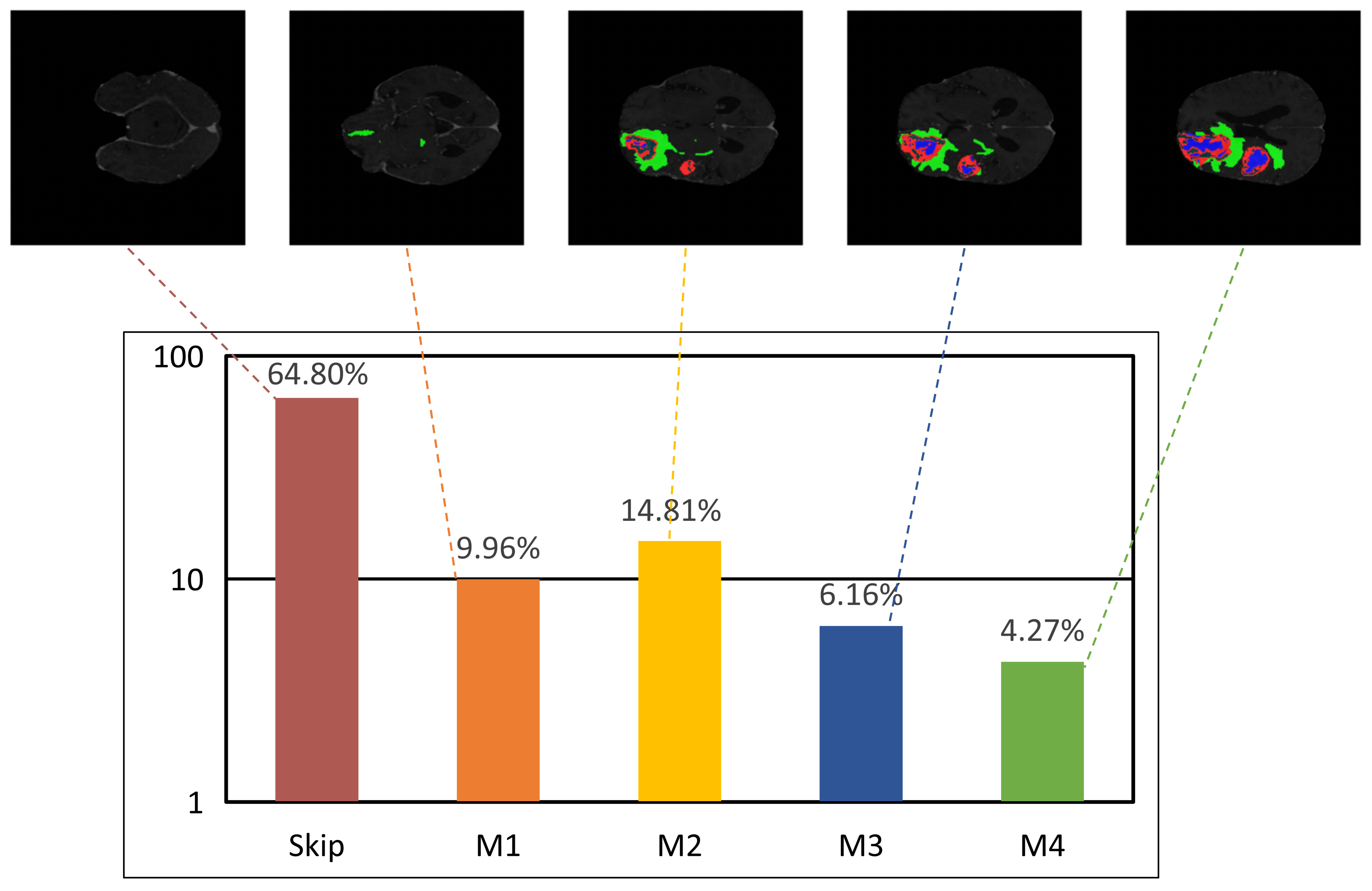}
    \caption{The activation ratio of each candidate model for different medical image slices in BraTS 2020 dataset. Skip, M1, M2, M3, M4 denote the operation of directly skip, candidate 1, candidate 2, candidate 3, and candidate 4, respectively.}
    \label{fig_supplementary_activation_brats2020}
    \vspace{-5pt}
\end{figure}

We further illustrate the activation ratio (or selection ratio) of each candidate model in the model bank for 3D volumentric segmentation in a slice-by-slice manner. Fig.~\ref{fig_supplementary_activation_brats2019} and Fig.~\ref{fig_supplementary_activation_brats2020} show the activation ratio of each candidate model on BraTS 2019 and BraTS 2020 dataset, respectively. 
During each inference, a single candidate in the Model Bank is activated according to the segmentation difficulty of the current slice. We can observe that the direct skipping operation accounts for a large portion for the MRI slices (i.e. more than half of the total number of slices). Moreover, since there are relatively more simple slices (e.g. those only contain one or two types of tumors and with small tumor regions) than the difficult slices, lightweight models (i.e. M1 and M2) are activated more over the large models such as M3 and M4. 
Through the proposed dynamic selection mechanism, an highly efficient and powerful architecture is achieved by our Med-DANet to reach a good balance between accuracy and computational efficiency.

\subsection{More Visual Comparison for Brain Tumor Segmentation}
\label{appendix3}

\begin{figure}[!tp]
    \centering
    \vspace{-5pt}
    \includegraphics[width=0.90\textwidth]{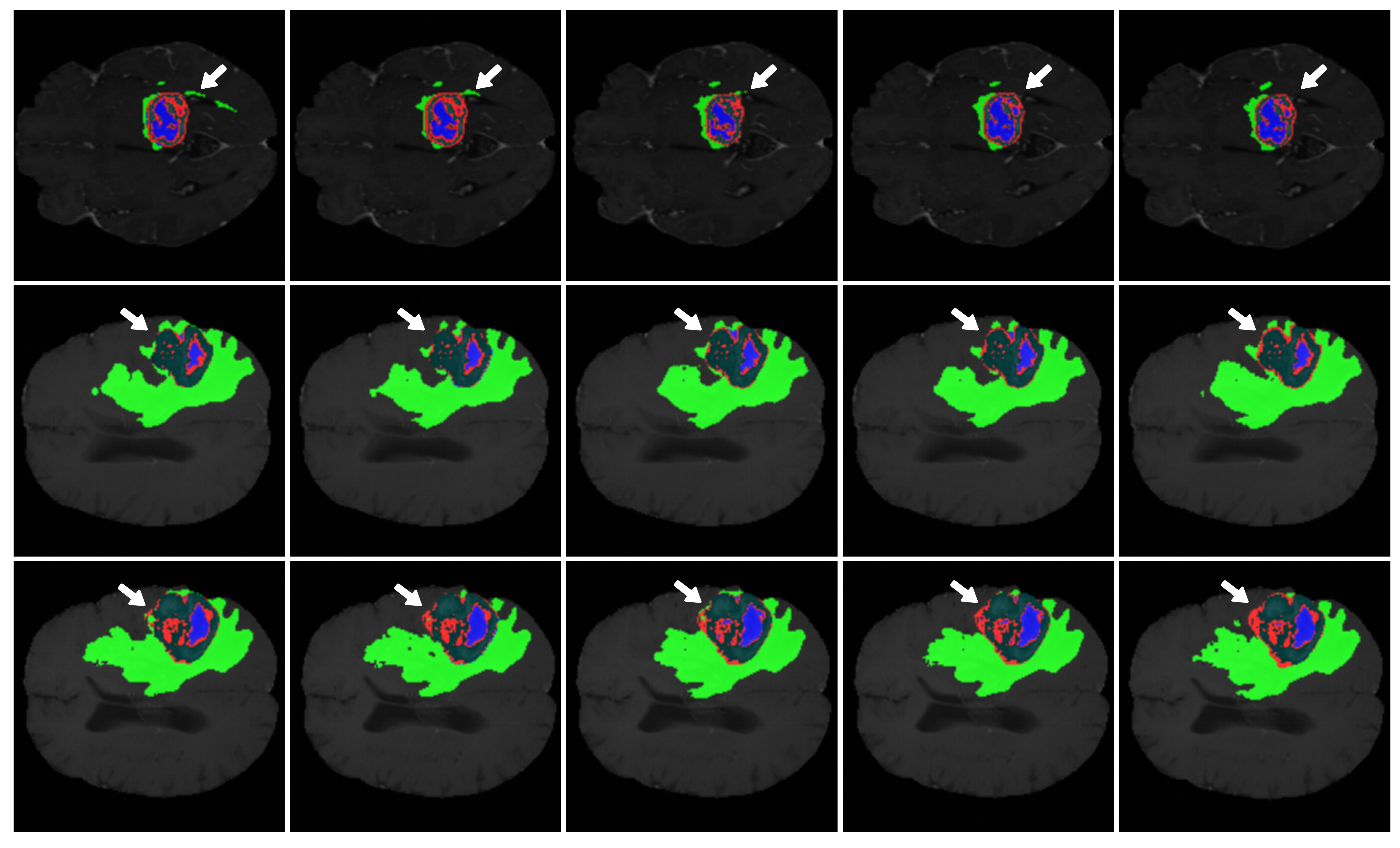}
    \begin{tabu} to 0.90\linewidth{X[1.0c] X[1.0c] X[1.0c] X[1.0c] X[1.0c]} 
        \scriptsize{3D U-Net} &  \scriptsize{VNet} &  \scriptsize{Att. U-Net} &  \scriptsize{\textbf{Ours}} &  \scriptsize{Ground Truth} \\
    \end{tabu}
    \caption{More visual comparison of MRI brain tumor segmentation results on BraTS 2019. The \textcolor{blue}{blue} regions denote the enhancing tumors, the \textcolor{red}{red} regions denote the non-enhancing tumors, and the \textcolor{green}{green} ones denote the peritumoral edema.}
    \label{fig_supplementary_brats2019}
\end{figure}

To further demonstrate the advantage of our proposed dynamic framework, we present more visualization of brain tumor segmentation results on BraTS 2019 for qualitative analysis in Fig.~\ref{fig_supplementary_brats2019}. The different methods utilized for visual comparison consist of 3D U-Net\cite{3dunet}, V-Net\cite{vnet}, Attention U-Net \cite{oktay2018attention}, and our Med-DANet. It is clear from Fig.~\ref{fig_supplementary_brats2019} that our framework can segment different kinds of brain tumors more precisely and generate much better fine-grained segmentation masks.

\clearpage

\bibliographystyle{splncs04}
\bibliography{reference}

\end{document}